\documentclass{article}
\usepackage{epsf}
\begin{document}
\baselineskip 12pt
\title
{Final State Interactions in Kaon Decays Revisited}

\author{M. Uehara\footnote{E-mail: ueharam@cc.saga-u.ac.jp}\\
Takagise-Nishi 2-10-17, Saga 840-0921, Japan}
\date{\today}
\maketitle
\begin{abstract}
We examine effects of final state interactions (FSI) in kaon decays using  
a new scheme without the Omn\'es function, and show that while 
$\Delta I=1/2$ rule is impossible to be explained in terms of FSI alone, 
$\epsilon'/\epsilon\sim 16.5\times 10^{-4}$ is obtained through FSI, if 
the calculation is concerned only with the CP-violating parts.  
The $\pi\pi$ contribution to the $m_L-m_S$ mass difference is shown to be 
about 14\%, if the sizes of the CP conserving amplitudes $|A_0|$ and 
$|A_2|$ are correctly given.
 \end{abstract}
\def\beq{\begin{equation}}
　\def\eeq{\end{equation}}
\def\beqa{\begin{eqnarray}}
　\def\eeqa{\end{eqnarray}}
\def\ba{\begin{array}}
　\def\ea{\end{array}}
\def\noeq{\nonumber}
\def\mpi{m_\pi} \def\fp{f_\pi}
　\def\mK{m_K} \def\fK{f_K}　
\def\der{\partial} 
\def\im{{\rm Im}} \def\re{{\rm Re}}
\def\cp{\frac{\epsilon'}{\epsilon}}

It seems that the established theory to explain simultaneously both of 
the $\Delta I=1/2$ rule and the large $\epsilon'/\epsilon$ ratio 
in kaon decays has not yet been obtained \cite{Bertol00,Bertol02,Buras98}. 
In order to take a step toward resolving the issues much efforts have been 
made on the evaluation of hadronic matrix elements of quark 
operators \cite{Hambye,Bijnens,ChQMBer}. 
Among them  it is advocated that taking account of final 
state interactions (FSI) is crucial to enhance or suppress the decay 
amplitudes besides giving the strong phases 
\cite{PPL,PPNPB,PPS,Paschos99}.  
The final state interactions (FSI) in the kaon decays are usually described 
in terms of the Omn\'es function \cite{Omnes} as seen in 
Refs.\cite{Truong,PPL,PPNPB,PPS}. The FSI effect written in terms of the 
Omn\'es function is, however, involved with some ambiguities, inherent in 
dispersion integrals, coming from the dependence on a subtraction, 
 a cutoff and a multiplicative arbitrary polynomial 
function \cite{Buras, Colangelo,Bertol02}. 
We propose, therefore, a new scheme in this note, which makes the decay 
amplitude satisfy the final state interaction theorem without any 
dispersion relations. \\

In order to evaluate our scheme of FSI, 
we briefly see FSI in  production processes of $S$-wave dipion  
states such as $\gamma+\gamma\to \pi\pi$,  $\phi\to \gamma\pi\pi$,   
cascade decays of heavy quarkonia and others.  
Let us consider a production amplitude of an $S$-wave dipion state
with a mass $\sqrt{s}$, called an $f$-state: It is composed of a 
direct production term of the $f$-state given by tree diagrams, 
which we call the Born term and denote by $B_f$, and 
an indirect production term, in which at first an $i$-state, denoted by $B_i$,  
is directly produced and then  transformed into 
the final $f$-state through the $s$-channel loop and the $S$-wave scattering 
amplitude $T_{if}$. The loop integral should be  properly  
regularized, and it may, then, depend on the renormalization scale 
parameter or a cutoff parameter. 
We can write the production amplitude, $F_f(s)$ as 
\beq
F_f(s)=B_f(s)+\sum_iG_i(s)\cdot T_{if}(s),
\eeq
where $G_i(s)$ is the loop integral composed of the meson propagators 
in the $i$-th channel, whose imaginary part (discontinuity along the
 physical cut) is given as 
\beq
\im G_i(s)=-B_i(s)\rho_i(s)\theta(s-s_i), \label{imG}
\eeq
where  $\rho_i=p_i/(8\pi\sqrt{s})$ with $p_i$\,($s_i$) being the 
CM momentum (threshold energy squared) of the $i$-th channel.  
We can easily see that the above production amplitudes satisfy the 
final state interaction  theorem, 
\beq
\im F_f=-\sum_iF_i^*\,\rho_i\theta(s-s_i)\, T_{if} \label{imP}
\eeq
owing to Eq.(\ref{imG}), where we define the sign and normalizaton of 
$T_{ij}$ so as to satisfy $\im T_{ij}=-T^*_{ik}\rho_kT_{kj}$. 
We have found it very effective to use 
this  scheme as shown in Refs.\cite{Mennessier,OO,Hirenzaki, MU06, MU11}.  
If the Born term contains two vertices, the loop diagram is a triangle, 
the third  vertex of which is the starting vertex of $S$-wave 
$T_{if}$ scattering. The triangle loop integrals are used in  
the two-photon collision processes \cite{Mennessier},  and the
 radiative $\phi$ meson decay \cite{Achasov, Kumano}. In the case that 
the Born term is made of a single vertex, $G(s)$ is a two-point loop integral. 
If the off-shell momentum dependence of the Born term is 
rewritten  as the on-shell part and the remaining off-shell part  
having a form of $p^2-m^2$, the off-shell factor $p^2-m^2$ cancels one of 
the propagators, and the divergence is absorbed into counter terms  in the 
Lagrangian with higher chiral order. 
The on-shell part is written as a factorized form of the Born term and 
the chiral loop integral such as 
$G_i(s)=B_i(s)\cdot J_i(s)$ as seen in ChPT and in weak ChPT. 
Here $J_i(s)$ is the renormalization scale dependent chiral loop 
integral \cite{GL}, and $\im J_i(s)=-\rho_i(s)\theta(s-s_i)$.
The scheme respects  $s$-channel unitarity over the left-hand 
contributions, and this is shown to be rather valid in low energy dimeson   
scattering below 1 GeV or more \cite{OOBS,OOP,GNP,MU04}. Vector or 
heavy mesons are allowed to be exchanged in the loop, but we discard 
them at all, because their contributions are not significant for low mass 
dimeson states. This scheme has some advantages over the scheme using 
the Omn\'es function;  our scheme is applicable to inelastic multichannel 
reactions,  it does not use dispersion 
integrals over the whole range of the physical cut, and then it is free of 
the ambiguities coming from the subtraction and from asymptotic 
behaviors of the integrands, though it depends on the logarithm  of the 
renormalization scale $\nu$.
Thus,  probably this scheme is also useful for the $K\to\pi\pi$ decays.\\
 
In order to incorporate the above scheme into the kaon decay amplitudes, we 
have to know the Born amplitudes, where the decay vertex is regarded as  a 
point, and does not have any discontinuities along the physical cut, 
$s\,>\,4\mpi^2$. As $B_I$ we take the real amplitude for 
$K_0\to(\pi\pi)_I$ given as \cite{BEF,Harada}
\beqa
B_0&=&\frac{G_F}{\sqrt{2}}V_{ud}V^*_{us}\frac{X}{\sqrt{6}}\left[-z_1
+2z_2+3z_4+3z_6\frac{Y_6^0}{X}\right],\\
B_2&=&\frac{G_F}{\sqrt{2}}V_{ud}V^*_{us}\frac{X}{\sqrt{3}}
\left[z_1+z_2\right],
\eeqa
where we take the large $N_c$ limit for $<Q_i>_I$ and 
\beq
X=\sqrt{2}f_\pi(\mK^2-\mpi^2) \quad 
Y_6^0=-4\sqrt{2}(f_K-f_\pi)\left(\frac{\mK^2}{m_s+m_u}\right)^2.
\eeq
For the Wilson coefficients $z_i$'s we take the values of the leading 
order (LO) approximation obtained in Ref. \cite{Buchalla} at $\mu=1$ GeV 
with $\Lambda^{(4)}_{\overline{MS}}=435$ MeV as 
\beq
(z_1,\,z_2,\,z_4,\,z_6)=(-0.901,1.541,-.016,-0.018),
\eeq
which give  
\beqa
B_0&=&\left\{
\ba{cr}
9.33\times 10^{-8}\,{\rm GeV}& m_s+\bar m=175\,{\rm MeV},\\
9.61\times 10^{-8}\, {\rm GeV}& m_s+\bar m=150\,{\rm MeV},\\
10.33\times 10^{-8}\, {\rm GeV}& m_s+\bar m=120\,{\rm MeV},\\
\ea \right.\\
B_2&=&1.94\times 10^{-8}{\rm GeV}. 
\eeqa

The amplitudes of $O(G_Fp^4)$ include one-loop 
contributions, which is of $O(N_c^{-1})$. The $s$-channel pion loop term 
is written 
\beq
T^{(2)}_I(s)\cdot J_{\pi\pi}(s)\cdot B_I(s)
\eeq
evaluated at $s=\mK^2$ \cite{PPS,BPP}, where $T^{(2)}_I$ denotes the 
$O(p^2)$ amplitude for $S$-wave $\pi\pi\to\pi\pi$ scattering with 
the isospin $I$. This structure is the same as we encounter in the strong 
ChPT amplitudes, and implies that the off-shell momentum dependence of 
$X$ is the same as described above.. Thus, we rewrite the decay amplitudes 
as 
\beq
A_Ie^{i\delta_I}=\left.B_I[1+J_{\pi\pi}(s)T_I^{\pi\pi}(s)] \right|_{s=\mK^2},
\label{CPcons}
\eeq
where we replace $T^{(2)}$ by the full $\pi\pi$ scattering amplitude $T_{ij}$, 
which can be calculated through a unitarized ChPT.  Higher order corrections 
should be added to $B_I$, except for the $s$-channel loop terms, 
but $B_I$ is left unchanged here in order to know the pure FSI effect. 
This means that we include higher order terms specially in the $s$-channel 
in order for the amplitudes to satisfy {\it exact unitarity on the 
elastic physical cut}.
\begin{figure}[h!]
\begin{center}
 \epsfxsize=12cm
 \centerline{\epsfbox{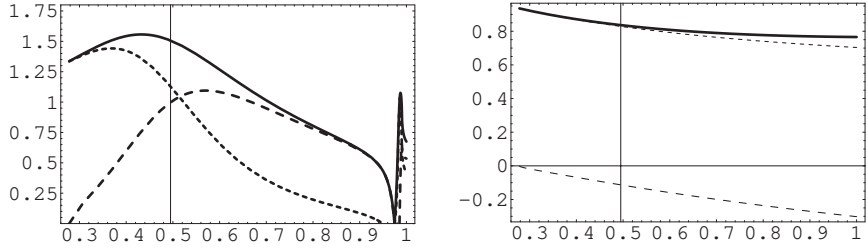}}
 \end{center}
 \label{fig:ff02}
 \caption{Energy dependence of $R_0(\sqrt{s})$ (left) and 
 $R_2(\sqrt{s})$ (right). 
 The solid lines are for the absolute values, and the dotted (dashed) lines for 
 the real (imaginary) parts. The vertical lines indicate the point at 
 $\sqrt{s}=m_K$.}
 \end{figure}
 The factor,
\beq
R_I(s)=1+J_{\pi\pi}(s)T_I(s)=|R_I(s)|e^{i\delta_I(s)},
\eeq
gives the whole effect of FSI coming from the strong $\pi\pi$ interaction, 
which is shown in Fig. 1.
 Using the scattering amplitudes calculated in terms of a unitarized 
 ChPT \cite{MU04}, we have  
\beqa
&&|R_0(\mK^2)|=1.506\mbox{  and  } |R_2(\mK^2)|=0.836,\\
&& \delta_0(\mK^2)=41.5^\circ\mbox{  and  }\delta_2(\mK^2)=-7.76^\circ,
\eeqa
where the renormalization scale parameter of  the 
$\pi\pi$ interaction is taken to be $\nu=1$ GeV.
We note that recent ChPT with dispersion relations gives 
$\delta_0(\mK^2)-\delta_2(\mK^2)=47.7^\circ\pm1.5^\circ$ 
\cite{CGL}.
We do not take into account of $K\bar K$ and $\eta\eta$ channels, because 
their effects are very small; for example, adding the $K\bar K$ loop integral 
$J_{K\bar K}(s)T_{K\bar K\to\pi\pi}(s)$ gives only 0.05 at $s=\mK^2$. We 
also note that it is shown that the contributions from the $u$-  and 
$t$-channel loop diagrams are small \cite{Colangelo}.

Thus, we obtain the values of the PC-conserving amplitudes except for 
the strong phase $\exp[{i\delta_0}]$ and $\exp[{i\delta_2}]$ as
\beq
A_0=(14.06,\, 14.48,\,15.56) \times 10^{-8}\,{\rm GeV},\, 
\mbox{ and  }\, A_2=1.62\times 10^{-8}\,{\rm GeV},
\eeq
where three values of $A_0$ correspond to the values of $m_s+\bar m$ at 
175 MeV, 150 MeV and 120 MeV, respectively. The size of $A_0$ is merely 
about a half of the experimental value $33.3\times 10^{-8}$ GeV, 
while the size of $A_2$ is rather close to the experimental 
value $A_2=1.50\times10^{-8}$ GeV. The ratio $\omega^{-1}=A_0/A_2$ is 
\beq
\omega^{-1}=8.68,\quad 8.94,\quad 9.60
\eeq 
depending on the value of $m_s+\bar m$, respectively, which is 
smaller than a half of the observed value 22.1.  
Since the enhancement factor by FSI 
would be almost the maximum, it is insufficient to explain the 
$\Delta I=1/2$ rule in terms of the FSI effects alone and then we have 
to search for higher order contributions or more fundamental mechanism 
to enlarge $B_0$ about twice.

The values of the enhancement of FSI are very similar to those obtained in 
Ref. \cite{PPS}, where the once-subtracted Omn\'es function is used with 
the parameter set, the subtraction point $s_0=0$ and the cutoff 
energy squared $\bar z=(1.6 ~{\rm GeV})^2$ of the 
dispersion integral. The dispersion integrals seem 
to be sensitive to these parameters, however.\\

Next, we consider the FSI effect on the $\epsilon'/\epsilon$ value. 
The ratio $\epsilon'/\epsilon $ is given as
\beq
\cp=e^{i\Phi}\frac{\omega}{\sqrt{2}|\epsilon|}\left[\frac{\im A_2}{\re A_2}
-\frac{\im A_0}{\re A_0}\right], \label{CPratio}
\eeq
where $\Phi=\pi/2-\delta_0+\delta_2-\theta_\epsilon\approx 0$, where 
$A_Ie^{i\delta_I}$ denotes here the full amplitude involving the 
CP-violating part. 
Usually, the CP conserving amplitudes,  $\re A_2$ and $\re A_0$, their ratio 
$\omega$ and  $|\epsilon|$ are set to their experimental values,  and 
the theoretical calculation is focused on the evaluation of the CP-violating 
amplitudes, $\im A_I$'s.  The ratio is estimated through the 
formula \cite{Jamin}
\beq
\cp\approx13\,\im\lambda_t\left[B^{(1/2)}_6(1-\Omega_{IB})-
0.4 B^{(3/2)}_8\right] \label{ratio}
\eeq
at $m_s(m_c)=130$ MeV, $m_t(m_t)=165$ GeV,
$\Lambda^{(4)}_{\overline{MS}}=340$ MeV and 
$\im\lambda_t=1.33\times 10^{-4}$. Taking both 
parameters $B^{(1/2)}_6$ and $B^{(3/2)}_8$ to be 1 at the large $N_c$ 
limit and $\Omega_{IB}=0.25$ in the lowest 
order estimate \cite{BG}, we have 
$\epsilon'/\epsilon\approx 6\times 10^{-4}$. 

The FSI effects modify these values to 
$B^{(1/2)}_6|R_0|=1.506$, $B^{(3/2)}|R_2|=0.836$ and 
$B^{(1/2)}_6\Omega_{IB}|R_2|=0.209$, since the last term is proportional to 
$\im A_2$ \cite{PPNPB}, and then the value of the square 
brackets in Eq.(\ref{ratio}) is 0.963, that is larger by a factor of 2.75 than 
the value at the large $N_c$ limit. Thus,  we  have 
\beq
\cp\approx 16.5 \times  10^{-4}.
\eeq
This enhanced ratio is almost the same as the values obtained in Refs.  
\cite{ PPL,PPNPB,PPS,Paschos99,Paschos01}. If higher order corrections 
except for the $s$-channel loops are taken into the calculation of
$B_6$ and $B_8$, we would have a larger value of $\epsilon'/\epsilon$. 
 
If we do not distinguish the physics underlying the CP-violating part from 
that of the CP-conserving part,  both numerator and denominator of 
the first and of the second term in the square brackets of 
Eq.(\ref{CPratio}) are multiplied by the same factor, $|R_2|$ and 
$|R_0|$,  respectively, and then the effect of FSI does not work at all. The 
parameter $\epsilon$ is also independent of FSI, but $\omega$ is suppressed.. 
In this case it is also impossible to understand  the large size of the direct 
CP violation in terms of the FSI effects.  \\

We discuss last the $\pi\pi$ contributions to the $K_L-K_S$ 
mass difference $\Delta m=m_L-m_S$. It is said that the short distance 
component dominates the mass difference through the effective 
$\Delta S=2$ Hamiltonian \cite{Buchalla, HN}. According to them 
typically 70\% of $\Delta m$ can be described by the short distance 
contribution, and the remaining of $\sim 30$\% in $\Delta m$ is attributed 
to the long distance component. Our interest is in the estimate of 
the $\pi\pi$ contribution to the mass difference.   
Pennington calculated this by using the once-subtracted dispersion relation 
with the experimental $\pi\pi$ S-wave phase shift, and  obtained  
$\Delta m/\Gamma_S=0.22\pm 0.03$ \cite{Pennington}, where 
$\Gamma_S$ is the total $K_S$ decay width and the experimental value 
is $7.36\times 10^{-15}\,{\rm GeV}$. 
 
We apply our scheme to the self-energy with $\Delta S=2$, 
and consider the following analytic function except for the physical cut; 
\beq
\Sigma(s)=-2\sum_{I=0,2}B_I(s)\left[J_{\pi\pi}(s)+
J_{\pi\pi}(s)T_I(s)J_{\pi\pi}(s)\right]B_I(s), \label{Sigma}
\eeq
which gives 
\beq
\Sigma(\mK^2)=2\mK\Delta m+i\mK\Gamma_S.
\eeq
Indeed, we have 
\beq
\im \Sigma(\mK^2)=\left.\rho_{\pi\pi}[2A_0^2+2A_2^2]\right|_{s=\mK^2}
=\mK\Gamma_S,
\eeq
where $A_I$ denotes the CP-conserving amplitude defined in 
Eq.(\ref{CPcons}).
We see that the energy dependence of $\im \Sigma(s)$  is almost the 
same below 1 GeV as the one by Pennington, where $\mK^2$ in $X$ of 
$B_I$ is replaced by $s$. 
\begin{figure}[h!]
\begin{center}
 \epsfysize=3cm
 \centerline{\epsfbox{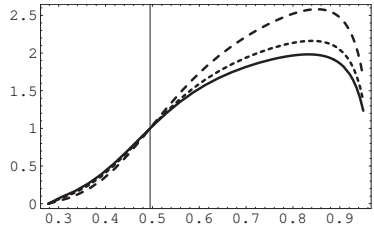}}
 \end{center}
 \label{fig:imag}
 \caption{Energy dependence of $\im \Sigma(\sqrt{s})/\im\Sigma(\mK)$. 
 The solid  (dotted, dashed) line is for $m_s+m_u=120\,(150,\,175)$ MeV. }
\end{figure}
The numerical value of $\Sigma(\mK^2)$ gives the ratio of 
$\Delta m_{\pi\pi}$ to $\Gamma_S$ as 
\beq
\frac{\re\Sigma}{2\,\im\Sigma}=\frac{\Delta m_{\pi\pi}}{\Gamma_S}=0.072\sim 0.074
\eeq
for $m_s+m_u=120$ and 175 MeV, respectively. 
We note that if we modify the sizes of $A_0$ 
and $A_2$ so as to reproduce the experimental ones, the above ratio does 
not change, but we have $\Delta m_{\pi\pi}/\Delta m^{\rm exp}=0.14$, 
which should be compared with the value obtained by Pennington, 
$\Delta m_{\pi\pi}/\Delta m^{\rm exp}=0.46$. \\

 It should be noticed that the enhancement by $R_0$ near the kaon mass is not 
due to the $\sigma$ resonance,  but due to the whole rescattering dynamics 
without the pre-existing $\sigma$ pole \cite{GM}. 
The effects expected by introducing the $\sigma $ pole through the 
linear $\sigma$ model \cite{Morozumi} or the 
nonlinear $\sigma$ model \cite{Harada}
can be replaced by the FSI effects without the degrees of freedom of 
the genuine $\sigma$ resonance. This replacement resolves a problem 
between the final state phase $\exp[i\delta_0]$ and the width ignored in the 
$\sigma$ propagator \cite{Morozumi}, and does the difficulty 
that the $\sigma$ pole dominance may give a too large contribution to the 
$K_L- K_S$  mass difference \cite{Terasaki}.

The scheme for the FSI effects explained in this note is useful in the 
sense that it is less worried by the ambiguities concerning with the 
dispersion integrals. We have shown in this note that FSI alone is impossible 
to reproduce the expected size of $A_0$, but it naturally gives a large  
$\epsilon'/\epsilon$ ratio, if we have a sufficient size of $|A_0|$. 
It turns out that the two-pion contribution to the $m_L-m_S$ mass 
difference is about  14 \% under the condition that the size of $A_0$ is 
sufficient. Thus, the crucial point for the $K_S\to \pi\pi$ decay is how
 to obtain $B_0\sim 22\times 10^{-8}$ GeV. 

\begin{center}
{\bf Acknowledgments}
\end{center}
The author thanks K. Funakubo, Saga University,  and K. Terasaki, 
Institute for Fundamental Physics, Kyoto University, for valuable 
comments. He also thanks the Department 
of Physics, Saga University for the hospitality extended to him. \\


\begin{thebibliography}{99}
\bibitem{Bertol00} S. Bertolini, hep-ph/0206095, Int. Workshop on 
Heavy Quarks and Leptons, Vietri sul Mare, Italy, 2002;
\bibitem{Bertol02} S. Bertolini, Proc. La Thuile 2000 [hep-ph/0007137].
\bibitem{Buras98} A.J. Buras, {\it Probing the Standard Model of 
Particle Interactions}, F. David and R. Gupta, eds, 1998 Elsevier Science B.V.
(hep-ph/9806471).
\bibitem{Hambye} T. Hambye, G.O. K\"ohler, and P.H. Soldan, Nucl. Phys. 
B564 (2000),391.
\bibitem{Bijnens} J. Bijnens and J. Prades, JHEP 9901 (1999), 023; {\it ibid.} 
0006 (2000),035.
\bibitem{ChQMBer} S. Bertolini, J.O. Eeg, M. Fabbrichesi and E.I. Lashin,
 Nucl. Phys. B514 (1998), 63.
\bibitem{PPL}  E. Pallante and A. Pich, Phys. Rev. Lett. 84 (2000),2568.
\bibitem{PPNPB}  E. Pallante and A. Pich, Nucl. Phys. B592 (2000),294.
\bibitem{PPS}  E. Pallante, A. Pich and I. Scimemi, 
Nucl. Phys. B617 (2001), 441.
\bibitem{Paschos99} E. A. Paschos, hep-ph/991223.
\bibitem{Omnes}  R. Omn\'es, Nuovo Cimento  8 (1958), 316.
\bibitem{Truong} T.N. Truong, Phys. Lett. B207 (1988), 495.
\bibitem{Buras} A.J. Buras, M. Ciuchini, E. Franco, G. Ishidori, G. Martinelli
and L. Silvestrini, Phys. Lett. B480 (2000), 80.
\bibitem{Colangelo} S. B\"uchler, G. Colangelo, J. Kambor and F. Orellana, 
Phys. Lett. B521 (2001), 22 and 29.
\bibitem{Mennessier}  G. Mennessier, Z. Phys. C16 (1983), 241.
\bibitem{OO} J. A. Oller and E. Oset, Nucl. Phys. A629 (1998), 739. 
\bibitem{Hirenzaki} E. Marco, S. Hirenzaki, E. Oset and H. Toki, 
Phys. Lett. B470 (1999) 20.
\bibitem{MU06} M. Uehara, hep-ph/0206141.
\bibitem{MU11} M. Uehara, Prog. Theor. Phys. 109 (2003), 265.
\bibitem{Achasov} N.N. Achasov and V.N. Iwanchenko, Nucl. Phys. 
B315 (1989),405.
\bibitem{Kumano} F.E. Close, N. Isgur and S. Kumano, Nucl. Phys. 
B389 (1993), 513.
\bibitem{GL} J. Gasser and H. Leutwyler, Nucl. Phys. B250 (1985), 465.
\bibitem{OOBS} J.A. Oller and  E. Oset, Nucl. Phys. A620 (1997), 438.
\bibitem{OOP}  J.A. Oller, E. Oset and J.R. Pel\'aez, 
Phys. Rev. D59 (1999), 074001, Errata ibid. D60 (1999), 09906 and  
D62 (2000), 114017.
\bibitem{GNP} A. Gom\'s Nicola and J.R. Pela\`ez, Phys. Rev. 
D65 (2002), 054009.
\bibitem{MU04} M. Uehara, hep-ph/0204020.
\bibitem{BEF} S. Bertolini, J. O. Eeg and M. Fabbrichesi, Rev. Mod. Phys. 
72 (2000), 65.
\bibitem{Harada} M. Harada, Y. Keum, Y. Kiyo, T. Morozumi, T. Onogi, N. 
Yamada, Phys. Rev. D62 (2000), 014002.
\bibitem{Buchalla} G. Buchalla, A.J. Buras and M.E. Lautenbacher, 
Rev. Mod. Phys. 68 (1996),1125.
\bibitem{BPP} J. Bijnens, E. Pallante and J. Prades, 
Nucl. Phys, B521 (1998), 305.
\bibitem{CGL}  G. Colangero, J. Gasser and H. Leutwyler, 
Nucl. Phys. B603 (2001), 125.
\bibitem{Jamin} M. Jamin, hep-ph/9911390, talk given at Heavy Flavour 
8, Southampton UK (1999).
\bibitem{BG} A.J. Buras and J.-M. G\'erard, Phys. Lett. B192 (1987), 156. 
J.F. Donoghue, E, Golowich, B.R. Holstein and J. Trampetic, Phys. Lett. 
B179 (1986), 361. 
\bibitem{Paschos01} E.A. Paschos, hep-ph/0109284, Invited talk at 
the Int. Conf. on CP Violation KAON 2001 (2001), Pisa, Italy.
\bibitem{HN} S. Herrlich and U. Nierste, Nucl. Phys. B419 (1994), 292; 
Phys. Rev. D52 (1995), 1505.
\bibitem{Pennington} M.R. Pennington, Phys. Lett. 153B (1985), 439.
\bibitem{GM} Ulf-G. Meissner, Comm. Nucl. Part. Phys. 20 (1991), 119. 
 S. Gardner and Ulf-G. Meissner, Phys. Rev. D65 (2002), 094004. See also 
 Ref.\cite{MU04}.
\bibitem{Morozumi} T. Morozumi, C.S. Lim and A.I. Sanda, Phys. Rev. 
Lett. 65 (1990), 404.
\bibitem{Terasaki} K. Terasaki, hep-ph/0008225, talk given at Workshop  
 on $\sigma$-Meson and Hadron Physics, Kyoto (2000). 


\end{thebibliography}
\end{document}